\documentclass{article}
\usepackage{amssymb}

\usepackage{graphicx}
\usepackage{amsmath}

\begin{document}

\hfill{ DSF-41/2000} \vskip2cm \baselineskip=16pt

\begin{center}
{\large \textbf{{Tunnelling Effects in a Brane System and Quantum Hall
Physics}\footnote{Work partially supported by the E C RTN programme
HPRN-CT-2000-00131} }}
\vskip1cm

Luigi Cappiello, Gerardo Cristofano, Giuseppe Maiella, Vincenzo Marotta,

\vspace{0.2cm}

{\small \ Dipartimento di Scienze Fisiche, Universit\`{a} di Napoli \\[0pt]
and INFN - Sezione di Napoli \\[0pt]
Via Cintia - Compl.\ Universitario M. Sant'Angelo - 80126 Napoli, Italy }

\end{center}

\vspace{1cm}

\noindent \textbf{PACS numbers}: $11.27 ; 73.40.Hm$

\noindent \textbf{Keywords}: String Theory, Quantum Hall Effect

\vspace{0.5cm}

\begin{quote}
{\small \textbf{Abstract} } We argue that a system of interacting D-branes,
generalizing a recent proposal, can be modelled as a Quantum Hall fluid. We
show that tachyon condensation in such a system is equivalent to one
particle tunnelling. In a conformal field theory effective description, that
induces a transition from a theory with central charge $c=2$ to a theory
with $c=3/2$, with a corresponding symmetry enhancement.
\end{quote}

\vfill
{\small
\begin{tabbing}
\\ E:mail:
\=cappiello(cristofano;maiella;vincenzo.marotta)@na.infn.it
\end{tabbing}}
\newpage

\section{\protect\bigskip Introduction}

\bigskip

Recently a pioneering paper has appeared \cite{Brodie:2000yz}, proposing a
strict analogy between non perturbative string phenomena and Quantum Hall
(QH) physics. The model is based on a particular configuration of D-branes
of type II A superstring theory compactified in three space dimensions. It
consists of a gas of D0 branes on a spherical D2 brane immersed in the
background magnetic field generated by a D6 brane at the center of the
sphere. It has been argued that the various interactions among the D-branes
lead to a (stable) topological configuration, called the Hall soliton. In
particular the system made of D0-D2 branes behaves as an incompressible Hall
fluid. It is our strong belief that such an analogy can shed light on old
and new phenomena in the Quantum Hall Effect (QHE) (see for instance \cite
{Prange}) and on non perturbative string effects, through the unifying
description provided by conformal field theory (CFT). In fact two
dimensional CFT methods have already been crucial in the world-sheet
description of strings \cite{FMS} and in an effective description of a QH
fluid \cite{lufub}, \cite{cgm2} at Jain hierarchical fillings \cite{J}, \cite
{CMM} and also non standard ones ($\nu =m/pm+2$) \cite{CMP}.

The aim of the paper is to extend the picture of ref.\cite{Brodie:2000yz}
by considering a system of two D2 branes carrying D0 branes on their
surface. For that system we argue that tachyon condensation due to strings
between the D0 and D2 branes takes places and produces a decrease of the
central charge of the (twisted) CFT describing this system from the value
$c=2$ to $ c=3/2$, corresponding to a new stable vacuum. In this process
some degrees of freedom drop out, carrying with them a $c=1/2$ contribution
to the original value of the central charge. The fields of the final CFT
realize, in a special case, a representation of superconformal algebra,
giving the raison d'etre of its stability.

The role of the tachyon in the D-brane era of string theories has been
recognized soon after the work of \ Polchinski \cite{Polchinski:1995mt}.
Both in the case of D-brane-anti-D-brane system \cite{Banks:1995ch} and in
the interaction between D-branes \cite{Polchinski:1996na}, which do not form
a\ BPS state, there is an instability. In the closed string channel it is
due to a non compensation between the forces due to the exchange of fields
of the NS-NS sector and those of the R-R sector; in the dual open string
channel, it corresponds to a wrong GSO projection on the spectrum of the
open strings stretching between the branes, giving rise to a tachyonic
ground state. The picture changes dramatically if the tachyon field develops
a non trivial potential and condensation takes place.

Such an idea originally appeared in the study of bound states of parallel
D-branes of different dimensions \cite{Witten:1996im}, \cite{Douglas:1995bn}%
. Successively it was shown that the exceeding energy produced in the
formation of a D(p+2)-Dp-branes bound state is compensated by the (negative)
value of the tachyon potential, which was evaluated in string perturbation
theory \cite{Gava:1997jt}. More recently Sen \cite{Sen:1998rg} conjectured
that in a D-brane-anti-D-brane system of type II superstrings the negative
contribution to the energy density coming from the tachyon potential
completely cancels the sum of the tensions of the two branes, giving rise to
a configuration with zero energy density equivalent to a closed string
vacuum and restoring full space-time supersymmetry. Since then a growing
amount of evidence has been presented in favour of it (see for instance
refs. \cite{SFT}, \cite{NC}, \cite{HARVEY}).

The plan of the paper is as follows. In Section 2, after a short review of
the brane configuration proposed in \cite{Brodie:2000yz} to describe the
two dimensional electron and magnetic flux system of a QH fluid, we show
that the vertex operator formalism allows for the CFT description of the
collective excitations, which are relevant both for the brane and Hall
physics. In Section 3 a system of two D2-branes is considered and a
tunnelling effect due to tachyon condensation is analyzed. As a result we
show that the central charge of the CFT \cite{CMP} decreases from the value
$ c=2$ to $c=3/2$, with a corresponding enhancement of the chiral algebra,
which for a particular value of the filling realizes an $\mathcal{N}=2$
super CFT. In Section 4 we present some  considerations, which motivated
our paper, emphasizing the new aspects of tachyon condensation mechanism
mutuated from the QHE analogy. Open problems as well as possible directions
for future work are also indicated.

\section{The physical picture}

Following ref.\cite{Brodie:2000yz}, let us first consider a D2-brane wrapped
around a cylinder with $K$ coinciding D6-branes placed along the cylinder
axis. All other extra dimensions are compactified. Although in superstring
theory it would be consistent to put two NS 5-branes at the two cylinder
edges, we will not analyze such setting here, i.e. we will consider an
infinitely long cylinder, so that some of the physical quantities which we
introduce in the following are to be understood as densities. The $K$
D6-branes act as a string of monopoles and their magnetic field couples,
through a Chern-Simons interaction term, to the worldvolume U(1) gauge field
$A$ of the $D2$-brane. As a result there is an induced electric background
charge on the D2-brane given by
\begin{equation}
Q=K
\end{equation}
In order to make the D2 electrically neutral, one must add $K$ fundamental
strings stretched between the D6 and the D2-brane, whose endpoints are
effectively seen as charges in the D2-brane worldvolume theory.\emph{\ }At
this point $N$ delocalised D0-branes must be disposed on the D2-branes,
where they appear as a nonvanishing magnetic flux, to energetically
stabilize the system. Indeed it is known that D0 and D6-branes repel each
other, so that they can counterbalance the attractive force between D6 and
D2 brane due to the tension of the stretched strings, which would otherwise
make the system collapse. The D0-branes produce N units of magnetic flux on
the D2-brane
\begin{equation}
\frac{1}{2\pi }\int_{D_{2}}F=N  \label{flux}
\end{equation}
where $F=dA$, is the field strength of the D2-brane gauge field.

The D0-branes then behave as an incompressible fluid \cite{Brodie:2000yz},
that is as a Quantum Hall fluid with filling
\begin{equation}
\nu =\frac{K}{N}
\end{equation}
If for example we take $K=N_e$ and $N=(2p+1)N_e$ the whole system, \emph{in
the large $N_e$ limit}, is stable.

It is well known \cite{lufub}, \cite{cgm2} that the ground state wave
function of the Hall system, at filling $\nu =1/2p+1$, can be described in
terms of correlators of vertex operators $V(z)$ given by

\begin{equation}
V(z)=e^{\frac{l}{R}Q(z)},\quad l=1,...,2p+1.  \label{VERTEX}
\end{equation}
where $Q(z)$ is a chiral scalar Fubini field, compactified on a circle of
radius $R$, with $R^{2}=1/\nu=2p+1$, and having the standard mode expansion
\begin{equation}
Q(z)=q-i\, p\, ln z + \sum_{n\neq 0}\frac{a_n}{n}z^{-n} \label{scalar}
\end{equation}
with $a_n$, $q$ and $p$ satisfying the commutation relations $\left[%
a_n,a_{n^{\prime}}\right]=n\delta_{n+n^{\prime},0}$ and $\left[q,p\right]=i
$.

The vertex operators (\ref{VERTEX}) are the primary fields of a CFT with
central charge $c=1$ and energy momentum tensor
\begin{equation}
T=-\frac{1}{2}(\partial_zQ)^2.
\end{equation}
The (analytic part of the) Laughlin ground state wave function \cite
{Laughlin} can be written as
\begin{equation}
<0|V_{2p+1}(z_{1})...V_{2p+1}(z_{n})|0>=\prod_{i<j=1}^{N}(z_{i}-z_{j})^{2p+1}
\end{equation}

It is interesting to notice that the vertex operators (\ref{VERTEX}) fully
describe the boundary states in a cylinder geometry \cite{cgm2}. In fact, if
we glue the cylinder boundaries, so obtaining a torus topology, the ground
state wave function of the Hall fluid is degenerate, with degeneracy equal
to $2p+1$, which is the number of the primary fields of the theory.
Pictorially, any given (internal) primary field propagates in (proper) time,
interacting with external electrons.

We point out that such a dyonic formulation of Laughlin wave function has a
remarkable interpretation in the string picture. The $z_{i}$ represent the
positions, on the $D2$-brane, of the end points of the fundamental strings
between the $ D2$ and the $D6$; they appear as pointlike charges immersed
in a background electric charge density of opposite sign. The $D0$-branes
present on the $D2$%
-brane feel the attractive forces of the electric charges, so that the
string end points and the D0-branes, surrounding them, tend to form
collective states, each having one unit of electric charge and $2p+1$ units
of magnetic flux. As a result the D2-D0 system behaves as a dyonic brane,
in perfect analogy to the condensate of a QH fluid at fillings $\nu=1/2p+1$
\cite{Laughlin}, \cite{cgm2}. Although a full field-theoretical treatment
of the formation of such collective states would be required, it is
noteworthy that this picture is supported by the analysis of the stability
of the Hall soliton made in \cite{BENA}. Further analysis of the dynamics
of the string endpoints on the worldvolume of the D2-brane  indicates that
the system actually has a quantum Hall behaviour for fractional filling
factors not too small ($\nu
\thickapprox 1/3$) \cite{Gubser:2000dz}.
 Using the  analogy between the two systems as
traced above, the collective excitation of one string endpoints and
D0-branes, surrounding it, is thus naturally described by the vertex
operator $V(z_{i})$ in eq.(\ref{VERTEX}). In the next section we shall
apply this effective description
 to study a setting of the Hall soliton model
containing  two  D2-branes.

\section{Tachyon condensation and tunnelling effects}

Let us now consider a configuration of two  coaxial cylindrical  branes, D2
and D2', with $K$ D6-branes still sitting along the axis of the cylinder.
 The $ K $ D6-branes induce a background charge on both D2-branes and
fundamental strings can be attached to them. The set of fundamental strings
is now richer, since it contains also strings stretching between D2 and
D2'. When  the two D2-branes are superimposed to form a bound state there
is an enhancement of the symmetry and  the D2-brane world-volume theory is
described by a $U(2)$ (Super) Yang-Mills gauge theory. Indeed, short
strings between D2 and D2'  can be identified with the non diagonal sector
of $U(2)$. (Delocalised) D0-branes, bound to the D2-branes, are represented
by a non vanishing magnetic flux given by eq. (\ref{flux}). More generally
when $m$ D2-branes form a bound state the integer value  $N$ of the flux
classifies the $U(m)$ bundles, and using the decomposition
$U(m)=(U(1){{\times}}SU(m))/Z_{m}$, which corresponds to separating the center of
mass (charged sector) from the remaining degrees of freedom (neutral
sector), standard arguments \cite{THOOFT} show that the neutral sector must
have a $Z_{m}$ twist, and  there is just one  current carrying charge and
not $m$.

 For $m$ D2 branes we consider  a  realization of an abelian orbifold CFT
in terms of $m$ scalar fields with central charge $c=m$, by an inductive
procedure acting on a $c=1$ CFT with a single scalar field \cite{CMM}. The
$m$ scalars can be organized into  an untwisted (charged) field and $m-1$
twisted (neutral) ones. Such a construction  has been successively applied
to  the analysis of paired and parafermionic states of a Quantum Hall
Fluid, with non standard fillings $\nu=m/(pm+2)$, in \cite{CMP}. The system
of two $D2$-branes we are considering, forming a (quasi-)bound state is
expected to be described by a  $c=2$ effective CFT of paired states. More
explicitly, starting from  a single chiral boson $Q(z)$, given in
 (\ref{scalar}),  we get the  untwisted field
\begin{equation}
X(z)=\frac{1}{2}\left( Q(z)+Q(-z)\right) ,
\end{equation}
compactified on a circle of radius $R^{2}=2/m=1$, representing the charged
sector, and a twisted field
\begin{equation}
\phi (z)=Q(z)-X(z)=\frac{1}{2}\left( Q(z)-Q(-z)\right)
\end{equation}
which satisfies the boundary conditions
\begin{equation}
\phi (e^{\pi i}z)=-\phi (z)  \label{eq: shift}
\end{equation}
and describes the neutral sector (with no zero-mode). Correspondingly, the
Virasoro generator is split in two terms \cite{VM2} both contributing with
$ c=1$ to the central charge. They are:
\begin{equation}
T_{X}(z)=-{\frac{1}{2}}\left( \partial _{z}X(z)\right) ^{2}  \label{STRESSX}
\end{equation}
and
\begin{equation}
T_{\phi }(z)=-\frac{\partial _{z}\phi (z)^{2}}{4}+\frac{1}{16z^{2}}
\label{STRESSFI}
\end{equation}
Notice the second term in (\ref{STRESSFI}), which is typical of twisted
scalar fields \cite{Ginsparg}. The primary fields $V(z)$ of the theory are
then  written as composite operators
\begin{equation}
V(z)=\mathcal{U}^{(\alpha)}_{c}(z)\psi (z),
\end{equation}
where the $\mathcal{U}^{(\alpha)}_{c}(z)=\frac{1}{\sqrt{z}}e^{i\alpha
X(z)}$ with $\alpha ^{2}=2$, while $\psi (z)$, the neutral sector, is built
out from the twisted scalar fields as
$\psi(z)=\frac{1}{\sqrt{z}}e^{i\sqrt{2}\phi(z)}$ \cite{CMP}. Moreover, the
highest weight states (h.w.s.) of the neutral sector contain two types of
chiral operators:  one which does not change the boundary conditions
\begin{equation}
\psi _{s}(z)=\frac{1}{2\sqrt{z}}\left( e^{i\sqrt{2}\phi (z)}+e^{i\sqrt{2}%
\phi (-z)}\right)   \label{even}
\end{equation}
and the other which does,
\begin{equation}
\psi _{a}(z)=\frac{1}{2\sqrt{z}}\left( e^{i\sqrt{2}\phi (z)}-e^{i\sqrt{2}%
\phi (-z)}\right) .  \label{odd}
\end{equation}
In the fermionized version one can see that they correspond to $c=1/2$
Majorana fermions, with periodic (Ramond) or anti-periodic (Neveu-Schwarz)
boundary conditions \cite{CMNP}. The corresponding bosonized energy-momentum
tensors are
\begin{equation}
T_{\psi _{s}}(z)=-\frac{1}{8}(\partial \phi )^{2}-\frac{1}{32z^{2}}\cos (2
\sqrt{2}\phi )+\frac{1}{32z^{2}}  \label{TSYMM}
\end{equation}
and
\begin{equation}
T_{\psi _{a}}(z)=-\frac{1}{8}(\partial \phi )^{2}+\frac{1}{32z^{2}}\cos (2
\sqrt{2}\phi )+\frac{1}{32z^{2}}.
\end{equation}
The description given above can be easily generalized to the entire series
with filling $\nu=2/2p+2$, which is equivalent to add $2p$ elementary
fluxes (D0-branes) to the system. The corresponding CFT has still $c=2$ but
$\alpha $ has now the values $\alpha
_{l}=l/2p+2$, with $l=1,...,2(2p+2)$.

Returning to the brane setting, we now argue that there is a tunnelling
phenomenon due to D0-branes which can migrate from one D2-brane to the
other. To better understand the following  physical picture, it may be
useful to recall  the role played by the tachyon in the formation of a
D0-D2 bound state. It is well known \cite{Polchinski:1996na}  that  the
original (non BPS) configuration of a D0-brane superimposed to a D2-brane
relaxes to a (BPS) state of minimal energy consisting of the D2-brane with
one unit of magnetic flux. Note that the transition from a non BPS state,
represented by a long supermultiplet of the space-time supersymmetry, to a
BPS state, which is a short multiplet, implies the disappearence \
(trasformation) of some of the degrees of freedom. As we mentioned in the
introduction, the binding energy can be ascribed to the  tachyon potential
which generates a new minimum of lower energy \cite{Gava:1997jt},
\cite{Sen:1998rg}. This mechanism is also responsible for the potential
barrier which prevents the D0-branes from escaping from the D2-brane in the
Hall soliton \cite{Brodie:2000yz}.

Let us now consider the two D2-branes with a small but finite separation. A
D0-brane located in between feels an actractive force due to the tachyonic
strings going from the D0 to any D2. Thus, the dynamical behaviour of the
D0-brane is determined by a non trivial tachyon potential with two minima,
symmetric with respect to the center of mass of the two D2-branes and
separated by a potential barrier which becomes lower when the D2-branes
distance decreases. In this configuration the  D0-brane undergoes a
tunnelling effect, and we expect that only symmetric states are selected
out. In our effective CFT description this is due to the cosine term in the
energy momentum tensor (\ref{TSYMM}), which produces a lowering of the
vacuum energy of the Ramond sector, {\it i.e.} of  the ($Z_2$ invariant)
degrees of freedom which survive the tunneling effect,  leading then to a
CFT with total central charge $c=3/2$.

The new  vacuum is annihilated by the total Virasoro generator
$L_0=L^X_0+L^{\psi}_0$, and for the special case of $p=1$, corresponding to
a filling $\nu=1/2$, it is also annihilated by the fermionic zero modes of
h.w.s. with conformal dimension $h=3/2$. In that case the resulting CFT is
actually  $\mathcal{N}=2$ superconformal theory, with the supercurrent
given by
\begin{equation}
J(z)=\partial X(z)+\frac{i}{\sqrt{2}}(\theta G^{-}(z)+\bar\theta
G^{+}(z))+\theta\bar\theta T(z)  \label{supercurrent}
\end{equation}
where $G^{{\pm}}(z)={\cal U}_c^{({\pm}2)}\psi_{s}(z)$, $T(z)=T_X(z)+ T_{\psi_s}(z)$
and $\theta$ and $\bar\theta$ are Grassmann variables.  That is a quite
important  result, since as a consequence of the tunnelling,  a new stable
D0-D2 bound state is formed, described by a superconformal theory, which
then guarantees the stability of the system. We consider that as a strong
support for  the consistency of our proposal.

 Although it is not the  aim of this paper to make  a detailed comparison
of our physical picture with the usual analysis of  tachyon condensation in
open strings  as a  boundary CFT (see  \cite{HARVEY} and references
therein),  some remarks are now in order. Let us consider a string scalar
field, $Y(\sigma)$ with $\sigma\in[0,\pi]$, interacting with a boundary
tachyon potential of the form
\begin{equation}
\lambda\cos (\beta Y(\sigma))|_{\sigma =0}, \label{tachpot}
\end{equation}
where $\lambda$  and $\beta$ are real parameters. Boundary CFT analysis
\cite{RG} shows that if the perturbation is relevant, as in the tachyon
case, the coupling constant $\lambda$ flows to an IR fixed point, pinning
$Y(0)$ to the values $Y_k=k\pi/\beta$ with integer $k$, corresponding to
the minima of the potential, which are to be identified with the positions
of D-branes along  $Y$. If we now consider $Y$ as the radial coordinate of
the open tachyonic strings of our model, we notice that the minima of the
tachyon potential have also a continous  degeneracy, which is due to the
global symmetries of the system, in fact,  there is no dependence on the
variables $z$, tangential to the D2-branes.  The $z$ are then continous
moduli parametrizing the vacua $Y(z)=Y_k$. In a low energy approximation,
the dynamics reduces to the motion  on the moduli space and we can add (at
lowest order in derivatives) a kinetic term of the form $(\partial
Y(z)/\partial z)^2$. In this way we can reconstruct  the effective CFT of
the neutral sector, which then plays a double role: on one side it
describes a boundary interaction for the fundamental strings ending on the
D2-branes, on the other side, it is a "bulk" effective theory on the D2
brane world volume.

As we have seen,  a cosine term of the form (\ref{tachpot}), with the
correct periodicity value, automatically appears in the energy-momentum
tensor (\ref{TSYMM}) of the (twisted) neutral sector.  In QH physics, such
an interaction term  goes under the name of interlayer one particle
tunnelling (see for instance \cite{PRYADKO} and references therein), and
$\lambda $ in (\ref{tachpot}) is the tunnelling amplitude.  Other attempts
to describe the transition from $c=2$ to $c=3/2$ have been developed  for
the QH fluid. In particular the models proposed in \cite{PRYADKO},
\cite{CABRA} share with our analysis  the relevance of tunneling processes
in determining such a transition, as well as the correct reduction of
degrees of freedom (i.e. the elimination of one Majorana fermion).

For $p>1$ the theory is not supersymmetric, but nonetheless it is invariant
under an enlarged symmetry, whose role, both in string theory and in the
QHE, is under study.

\section{Conclusions and outlook}

The Quantum Hall soliton introduced in \cite{Brodie:2000yz} is a powerful
source of inspiration for those who think that the two fields of string
theory and Quantum Hall physics have many contact points, through the
unifying description provided by CFT. In that framework the vertex
operators give a simple interpretation of the collective excitations
relevant to the physics of the QHE, both for Jain hierarchical fillings and
for non-standard ones. We have pushed forward the analogy between string
theory and the QHE by analyzing tachyon condensation in a non BPS system of
D-branes. In fact we find that this phenomenon is analogous to the
tunnelling between two layers in the QHE. The effective CFT description
given in \cite{CMP} fits very well in this setting and shows that tachyon
condensation leads to a new stable vacuum with higher symmetry.

The main novelty of the mechanism of tachyon condensation proposed here is
the role played by the magnetic flux (carried by the D0-branes), and the
setting of two (or more generally $ m $) D2-branes. As it has been
discussed, the description of the dyonic Laughlin particles in terms of
vertices for the Hall plateaus is crucial for understanding the analogy
between the brane setting and the QH system. Moreover,  for the setting of
at least two (or $m>2$) D2-branes one can give a description  in terms of
Fubini scalar fields (and corresponding vertices) with different boundary
conditions: the field with periodic b.c. (the charged sector) and the one
 (or $m-1$) with twisted $Z_2$ ($Z_m$) phases (the  neutral sector). Altogether they
build up the complete set of vertices (h.w.s) of the relevant CFT. In this
context the role of the  $Z_2$ ($Z_m$) symmetry acquires a peculiar
relevance; such a discrete symmetry couples the charged and the neutral
vertices, hinting to a more general description provided by  Matrix Theory
with $U(2)$ ($U(m)$) symmetry \cite{matrix}. Nevertheless, an interesting
result has already been found:  for the setting of two interacting
D2-branes we have obtained an effective description in terms of a
two-dimensional $\mathcal{N}=2$ super CFT. The generalization to the case
of $m>2$ should reveal new interesting aspects of the effective CFT
description of $m$ D2-brane setting.

\end{document}